\documentclass[preprint]{aastex}

\newcommand\beq{\begin{equation}}
\newcommand\eeq{\end{equation}}
\newcommand\beqar{\begin{eqnarray}}
\newcommand\eeqar{\end{eqnarray}}

\newcommand\ee[2]{#1 \times 10^{#2}}

\newcommand\pref[1]{(\ref{#1})}

\newcommand\etal{{\it et al.~}}

\newcommand\msol{\hbox{$M_{\odot}$}}
\newcommand\lesc{\ell_{\rm esc}}

\newcommand\snr[1]{\mbox{${\cal R}_{_{\rm #1}}$}}

\begin{document}

\title{DIFFUSE GAMMA-RAYS FROM LOCAL GROUP GALAXIES}

\author{Vasiliki Pavlidou \and Brian D. Fields}
\affil{Department of Astronomy and Center for Theoretical Astrophysics\\ 
University of Illinois\\
Urbana, IL 61801, USA}
            
\begin{abstract} 

Diffuse $\gamma$-ray radiation in galaxies is produced by
cosmic ray interactions with the
interstellar medium.
With the completion of EGRET observations, the only extragalactic
object from which there has been a positive detection of diffuse $\gamma$-ray
emission is the Large Magellanic Cloud.  
We systematically estimate the expected diffuse $\gamma$-ray flux
from Local Group galaxies, and determine
their detectability
by new generation $\gamma$-ray observatories such as GLAST. For each
galaxy, the expected $\gamma$-ray flux depends only on
its total gas content and its
cosmic ray flux.  
We present a method for calculating cosmic ray flux in these galaxies
in terms of the observed 
rate of supernova explosions, where cosmic ray acceleration is
believed to take place. The difficulty in  deriving accurate supernova
rates from observational data is a
dominant uncertainty in our calculations. 
We estimate the $\gamma$-ray flux for Local Group galaxies and find that
our predictions are
consistent with the observations for the LMC and with the observational
upper limits for the Small Magellanic Cloud and M31.  Both the Andromeda
galaxy, 
with a flux of $\sim 1.0 \times 10^{-8}$ photons 
sec$^{-1}$  cm$^{-2}$
above $100$ MeV, and the SMC, 
with a flux of $\sim 1.7 \times 10^{-8}$ photons 
sec$^{-1}$  cm$^{-2}$ above $100$ MeV, are 
expected to be observable by GLAST.  M33 is at 
the limit of detectability with a flux of 
$\sim 0.11 \times 10^{-8}$  sec$^{-1}$ cm$^{-2}$. 
Other Local Group galaxies are at least two 
orders of magnitude below GLAST sensitivity. 

\end{abstract}



\section{Introduction}
All-sky EGRET images (Hunter \etal 1997) 
dramatically show that the
$\gamma$-ray
flux above 100 MeV is dominated by emission from the Galactic disk.
This emission can be well understood (Strong 1996,  Sreekumar \etal\ 1998,
Hunter \etal\ 1997) in terms of cosmic ray interactions with the interstellar
medium. At energies 
 $\ga$  100 MeV, the generation of diffuse $\gamma-$ray emission is 
dominated by the decay of $\pi^0$ produced in collisions between cosmic
ray nuclei and interstellar medium nuclei. At lower energies, the dominant
emission mechanism is bremsstrahlung from energetic cosmic ray electrons.
Given the prominence of the Milky Way diffuse $\gamma$-ray emission, 
it is natural to ask whether we can see the corresponding emission
originating
from other galaxies.

To explore the possibility of such a detection requires an
understanding of the 
$\gamma$-ray emissivity due to interactions of cosmic rays with matter. 
The problem of calculating the $\gamma$-ray emissivity per hydrogen atom
given a cosmic-ray spectrum has been treated in detail by
Stecker  \cite{stecker1,stecker2,stecker3} and Dermer
\cite{dermer}, who have made use of both experimental data and
theoretical models on the relevant
collisional cross sections. The results of these studies are also
consistent with more
recent calculations by Mori \cite{mori} who used Monte Carlo event simulators
and recent accelerator data. These studies reproduce the Milky Way (MW)
$\gamma$-ray
flux quite well as a function of energy and angle on the sky
(at least in the 30--500 MeV range;
at higher energies, the situation is less
clear, and inverse Compton scattering of a hard electron
component may be important; see Strong, Moskalenko,
\& Reimer \cite{smr}). 

By combining this information with 
an estimate of the gas content and the cosmic
ray flux level and spectrum for a given galaxy, one can arrive at a
prediction for its total $\gamma$-ray flux.  Extended studies of this kind
have been made for the Magellanic Clouds.
Fichtel \etal \cite{fichtelmc} calculated the flux expected from the 
Large Magellanic Cloud (LMC). Their estimate of the cosmic ray flux was
based on
the assumption that the expansive
pressures of the gas, magnetic field and cosmic rays 
are in dynamic balance with the gravitational
attraction of matter.  Subsequent EGRET observations (Sreekumar {\it et
al} 1992) resulted in the detection of the LMC with a
$\gamma$-ray flux consistent with the prediction by Fichtel \etal
\cite{fichtelmc}. 

Sreekumar \& Fichtel \cite{sreeksmc} arrived at three different
predictions for the Small Magellanic Cloud (SMC) diffuse $\gamma$-ray
flux, each
one based at a different assumption for the level of the SMC cosmic ray
flux: a dynamic balance among thermal, magnetic, and cosmic ray pressures 
(as in the case of the LMC); a constant ratio of
cosmic ray electrons and protons (where the electron flux was deduced from
observational synchrotron data); and a ``universal'' cosmic ray flux (at the same
level as in the solar neighborhood). EGRET observations 
have not led to a positive detection of the SMC
in high-energy $\gamma$-rays, placing an upper limit to the $\gamma$-ray flux
which excludes the predictions based on a dynamic balance or a ``universal''
cosmic ray flux 
(Sreekumar \etal 1993, Lin \etal 1996). In this way, the debate over the
origin of cosmic rays
(Galactic versus extragalactic) was settled observationally and the cosmic
rays were shown to be originating from within the Galaxy.

Apart from the Magellanic Clouds, the only other Local Group galaxy
for which there have been theoretical $\gamma$-ray flux predictions is 
M31.\footnote{Theoretical studies have also been made for starburst
galaxies such as M82 (Aky\"{u}z \etal 1991) and NGC253 (Paglione \etal
1996) since these objects were good candidates for
detection by EGRET due to their high supernova
 rate and presumably high cosmic ray flux.
However, no positive detection of these objects has been achieved
(Blom \etal 1999).} 
\"{O}zel \& Berkhuijsen \cite{ozel} and \"{O}zel \&
Fichtel \cite{ozfichtel}, based on the observational data available at the time
for the distance and gas content of M31, concluded that if the cosmic ray
flux in the Andromeda galaxy is comparable to that in the Milky Way, then
the galaxy should be detectable by EGRET. However, 
Blom, Paglione, \& Carrami{\~n}ana \cite{blom} showed that EGRET 
has not detected M31, and have instead placed an upper limit for its
$\gamma$-ray flux lower than the theoretically  predicted value of
\"{O}zel \& Berkhuijsen \cite{ozel}. 

In this work we systematically study the $\gamma$-ray emission from
Local Group galaxies and its detectability;
with the completion of EGRET observations and the prospect of the
construction of new, more sensitive $\gamma$-ray observatories,
such an investigation is timely.
We survey the latest
available data for the gas contents of various Local Group
galaxies, with attention to the
uncertainties in the observational inputs,
and their impact on the predicted emission.
We also present a new method for computing the global mean cosmic ray 
flux and, thus the $\gamma$-ray flux,
from the observed properties of the extragalactic sources.
To do this we
use the ratio of the supernova (SN) rate in each galaxy to the
supernova rate of the Milky Way to calibrate the magnitude 
of the cosmic ray flux. This
association is justified by the fact that cosmic ray acceleration is
believed to take place in supernova remnants, an idea supported both by
theoretical arguments 
(e.g., Ellison, Drury, \& Meyer \cite{edm}) 
as well as observational evidence
(e.g., Koyama \etal 1995, Combi \etal 1998). 
Furthermore, we find this method leads to an estimate of
the LMC $\gamma$-ray flux which is in excellent agreement
with the level observed by EGRET.

Our predictions for the Local Group will be testable
by the forthcoming {\it Gamma-ray Large Area Space Telescope} (GLAST), 
which is predicted to be launched in 2005.  
The proposed design of GLAST, described in De Angelis \cite{deangelis}
and in the GLAST webpage ({\tt \url{http://glast.gsfc.nasa.gov}}), 
is that of a 1 ${\rm m^2}$ effective area detector, sensitive to 
energies from 20 MeV to 300 GeV, with a field
of view of 2.4 sr. This large field of view will allow any individual
source to be observed for 20\% of the duty cycle when GLAST is operating
in the normal sky-scanning mode. The single-photon angular resolution 
will be $3^\circ$ at 100 MeV and about $0^{\circ}.2$ at 10GeV. The nominal
predicted lifetime of the GLAST mission is 5 years, with a goal of 10
years of operation. These
specifications imply a sensitivity for GLAST of $\sim 2 \times 10^{-9} 
\,\,{\rm photons \,\,cm^{-2}\,s^{-1}}$ (5-$\sigma$ detection for a point
source at high Galactic latitudes after a 2-year all sky survey), better
by more than an order of magnitude than that of EGRET. 

Recent work by 
Digel, Moskalenko, Ormes, Sreekumar, \& Williamson \cite{dmosw}
has drawn attention to GLAST's potential for observing
the three brightest Local Group sources.
These authors presented simulated maps of the diffuse emission from
the Magellanic clouds 
using models of Sreekumar,
and discussed the observability of
M31 on the basis of current observational upper limits.
Our work gives theoretical support to this effort, 
as we make a robust prediction that M31 will be detected
by GLAST.  By considering the entire Local Group, we also
find that M33 could be detectable by GLAST.

In Section 2 we present the theoretical background of our calculations.
We discuss the $\gamma-$ray emissivity per H
atom for a given cosmic ray spectrum, by now
established both theoretically and observationally.
The ``leaky box'' model for cosmic
ray propagation leads to the
prediction that the CR flux should be proportional to a mean  SN rate in
the galaxy under consideration. Using this
fact and an estimate of the gas content of each galaxy of the Local Group,
we arrive at a prediction for its $\gamma$-ray flux.
In Section 3 we present the data on the SN rates of the MW and the
galaxies under study as well as their gas contents. In Section 4 we
describe our results, compare our predictions for the Magellanic Clouds
and
M31 with the EGRET observations, and compare these and predictions for
other Local Group galaxies with the anticipated sensitivity of GLAST. Discussion of
our findings and
conclusions follows in Section 5.  

\section{Gamma-Ray Production}
\label{sec:theory}

Inelastic collisions of high-energy cosmic ray protons with
the interstellar
medium protons results most frequently to the production of neutral pions
($\pi^0$) which then decay, with a probability $\sim$ 99\%, to two gamma
ray photons: 
\[
\begin{array}{lrll}
p+p \rightarrow&p+p\,+&\!\!\!\!\pi^0& \\ 
 & &\!\!\hookrightarrow&\gamma+\gamma
\end{array}
\]
If $\phi^{p}_E(T_{p})$ is the differential energy spectrum of the cosmic
ray protons (protons/(cm$^2$-s-GeV-sr)) and $\langle \zeta \,
\sigma_{\pi}(T_{p}) \rangle$
is the inclusive cross-section for the production of $\pi^0$ from p-p
collisions, then the total $\pi^0$ production rate per target H atom
($\Gamma_{\pi^0}$) is given
by:
\begin{equation}\label{dermeq1}
\Gamma_{pp \rightarrow \pi^0}=4\pi\int_0^{\infty} dT \, \langle \zeta \,
\sigma_{\pi}(T)\rangle 
\, \phi^{p}_E(T)
\end{equation}
In the case of the Milky Way, various authors 
(Stecker 1970, 1973, 1988;
Cavallo \& Gould 1971; Stephens \& Badhwar 1981;
Dermer 1986; Mori 1997) have calculated
$\Gamma_{\pi^0}$. The results are 
consistent with each other within errors. Here we adopt
\beq
\Gamma_{pp \rightarrow \pi^0} \approx 
7.0 \times 10^{-26} \,\,\pi^0
\left(\mbox{s - 
H atom }\right)^{-1}
  .
\eeq
There is an additional contribution to $\pi^0$ production by
interactions involving nuclei with $A>1$ either in cosmic rays or in the
interstellar
medium (predominantly $p + \alpha$ and $\alpha + \alpha$).
This contribution
increases the estimated pion production rate by a
multiplicative factor (Stecker 1970), which, when 
including $\alpha$ as well as nuclei heavier than helium
both in the cosmic rays and in the interstellar medium, is 
$\sim 1.5$ (e.g., Mori \cite{mori} and refs.\ therein). 
Taking into account this correction
and the fact that each pion produces two $\gamma$-rays, the 
{\it total}
$\gamma$-ray emissivity per hydrogen atom from {\it $\pi^0$ decay only}
is
\begin{equation}\label{allpi}
q_{\gamma}^{\pi^0}=2.0 \times 10^{-25} \,\,\, \mbox{photons} 
\left(\mbox{s - H atom}
\right)^{-1}
\, , 
\end{equation}
in good agreement with the theoretical calculations cited
above. Equation (\ref{allpi}) includes $\gamma$-rays of all
energies.
Now if we confine our interest in the energy range $>100$MeV, this
emissivity decreases  by a factor of 0.76. In the same energy range, Fichtel
 and Kniffen \cite{fichtel1}
have estimated the contribution in the $\gamma$-ray emissivity from cosmic
ray electron bremsstrahlung to be 55\% of the flux produced by neutral
pion decay. 
\footnote{The situation above 500 MeV is perhaps less clear. 
Strong, Moskalenko, \& Reimer \cite{smr} have recently
suggested that the dominant source in this regime
may be inverse Compton 
scattering of interstellar radiation off a hard electron component.}
In this way, we can calculate a total Galactic $\gamma$-ray
emissivity per hydrogen atom, for photon energies $ >100$ MeV ,
originating
by all $\pi^0$-producing CR-ISM collisions as well as bremsstrahlung
radiation from cosmic ray electrons:
\begin{equation}\label{emis}
q_{\gamma}(>100 \mbox{MeV}) = 2.4 \times 10^{-25} \,\,\, \mbox{photons /
(s - H
atom)}
.
\end{equation}
which is consistent with observational data
(e.g., Digel \etal 1995)

In order to extend this calculation to galaxies other than our own, 
we must relate the $\gamma$-ray production and hence cosmic ray flux,
to observable properties of the galaxies.
We therefore must account for both the acceleration and
propagation of cosmic rays, and their dependence on the galactic
environment.
The assumption that supernova explosions are the engines of CR
{\em acceleration} is encoded in simple and direct way. 
Specifically, we will impose 
a scaling of the CR source (injection) rate $Q_p$ with \snr{G}, the
mean SN rate in a specific galaxy $G$:
\[Q_{p}^{G} \propto \snr{G} .\]

To describe the {\em propagation} of cosmic rays requires more detailed treatment.
Cosmic rays propagate diffusively as 
they spiral along galactic magnetic field lines.
The particles sources are acceleration sites (i.e.,  supernova remnants) 
which are distributed
inhomogeneously; the sinks are energy losses due to 
a variety of processes, as well
as escape from the galaxy.
Cosmic ray propagation is thus properly described by a diffusion equation
which accounts for all of these effects as a function of position 
in the galaxy.  
However, for our purposes of describing a galactic-averaged cosmic ray flux, 
the treatment of propagation can be stripped to its essential features.
To do this, we make the following simplifying assumptions:
\begin{itemize}
\item The shape of the cosmic ray energy spectrum is the same throughout
each galaxy and identical to that in the Milky Way (although the
normalization may be different).
\item The ratios of $p$ to $\alpha$ as well as the ratio of CR electrons to
CR protons is constant and same to that in the Milky Way.
\item The propagation of cosmic rays in the galaxy can be approximated in
terms of the ``leaky box'' model.
\end{itemize}
The first two assumptions appeal to a universality in the
underlying physics of cosmic ray acceleration by individual supernova remnants.
The last assumption treats the galaxy as a single zone to be modeled, as we now see.

According to the ``leaky box'' model, the galaxy is treated as a homogeneous 
containment environment where CRs propagate freely with a certain probability 
per unit time to escape. Within the frame of this model the propagation 
equation becomes
\begin{equation}\label{fulllb}
\frac{\partial N_i (T,t)}{\partial t}=Q_i(T,t)+
\frac{\partial}{\partial T}
[b_i(T)N_i(T,t)]-\frac{1}{\tau_{\rm esc}}\,N_i(T,t) . 
\end{equation}
Here $N_i(T,t)$ is the density of particles of species i
 with kinetic energy between $T$ and
$T+dT$.  $Q_i(T,t)$ is the source term (particles per volume per time per
energy interval $dT$) including all sources of species $i$. 
As we are concerned with $p$ and $\alpha$ particles,
the source term $Q_i$ simply represents the supernova
acceleration as a source these ``primary'' species
(as opposed to ``secondary'' particles created in flight, such as Li, Be, and B). 
Additional spallation 
contribution and losses are negligible at our level of accuracy.
The second term in the right hand side of eq. (\ref{fulllb}) represents 
energy losses due to ionization of the ISM, with a rate 
$b_i(T)=-\left({\partial T}/{\partial t}\right)_i$;
in the energy range of 
interest for $\pi^0$ production ($> 279$ MeV), this can be considered unimportant
 to a good approximation. 
The loss of CRs that escape
from the ``leaky box'' is included in the last term, with $\tau_{\rm esc}$ 
being the mean time spent by the CRs in the containment volume.
Although strictly speaking inelastic collisional
losses should have also been included in the propagation equation, 
we have omitted them since 
the mean free path against collisions is much larger than 
the one against escape. 

If in addition we assume a steady state, the left hand side of eq.
(\ref{fulllb}) vanishes
  and the propagation equation for protons assumes the simple
form
\begin{equation}\label{simplelb}
0=Q_p(T)-\frac{1}{\tau_{esc}}N_p(T)\,.
\end{equation}
Physically, this corresponds to an equilibrium between sources (SN 
acceleration) and sinks (escape). 
Since the CR proton flux and the corresponding number
density are related via $\phi_p(T)=v_p N_p(T)$, 
we find that the (propagated) flux is given by
\begin{equation}\label{finallb}
\phi_p(T) = \lesc Q_p(T)    ,
\end{equation}
where $\lesc =\tau_{\rm esc}v$ is the mean free path against escape.\footnote{
It is often conventional to define
$q_p=Q_p/\rho_{_{\rm ISM}}$ and thus equation  (\ref{finallb}) implies
$\phi_p(T) = \Lambda q_p(T)$
$\Lambda = v_p \tau_{\rm esc}\,\rho_{_{\rm ISM}}$
is the escape pathlength in ${\rm g \ cm^{-2}}$, the
``grammage.''}

Thus, to make further progress in estimating the CR flux $\phi_p(T)$
 in the galaxy $G$, we need to have some understanding of the CR 
confinement in that 
galaxy, which enters in eq. (\ref{finallb}) through $\lesc$.
This depends on the details of the magnetic field
strength and configuration in these galaxies. 
The detailed physical origin of the confinement scale $\lesc$
(or equivalently, the diffusion tensor) is as yet uncertain,
but is almost certainly related to the structure of 
and fluctuations in the 
galaxy's magnetic field
(e.g., Berezinski\u{\i} et al.\ cite{bbdgp}).
Lacking any better knowledge, we will assume confinement conditions similar
to those in the Milky Way.  Specifically, we will
assume $\lesc$ is the same as in the Milky Way.
This amounts to an {\em Ansatz} that the physical properties that
determine $\lesc$ are dominated by local rather than global properties of
the host galaxy.
This assumption becomes more plausible the more
similar $G$ is to the MW, so we expect our approach to yield better 
results in the cases of M31 and M33 rather than in the cases of the 
Magellanic Clouds and other irregular galaxies.
(Alternatively, one could turn the problem around, and with $\gamma$-ray observations of
these objects, one can
measure or limit the cosmic ray confinement in these objects.)

Under this assumption, the CR flux is proportional to the SN rate in $G$:
\begin{equation}\label{def_fg}
\frac{\phi_{p}^G}{\phi_{p}^{MW}}=\frac{\snr{G}}
{\snr{MW}}=
f_{_{\rm G}}
\end{equation}
So from equations (\ref{dermeq1}) and (\ref{def_fg}) we get 
\begin{equation}\label{gammawithfg}
\Gamma_{\pi^0} ^G=f_{_{\rm G}}\Gamma_{\pi^0}^{MW}
\end{equation}
which, following the same procedure that has lead to eq. (\ref{emis}), finally 
gives
\begin{equation}\label{emiswithfg}
q_{\gamma}^G(>100 \mbox{MeV}) = 
2.36 \times 10^{-25}  f_{_G}\,\,\, \mbox{photons} \left(
\mbox{s - H
atom}\right)^{-1}
.
\end{equation}

Assuming that the CR flux level and spectrum remains the
same over the whole galaxy, this emissivity is space-independent.
Thus, the $\gamma-$ray flux from galaxy $G$ which lies 
at a distance $d$ and has a gas content of $M_{\rm gas}$ will simply be
\[F_{\gamma}^G=\frac{1}{ 4 \pi d^2}\,\frac{M_{\rm gas}}{m_p} \, 
q_{\gamma}^G \,,\]
or, using eq. (\ref{emiswithfg}) for $q_{\gamma}^G$,
\begin{equation}\label{flux}
F_{\gamma}^G(>100 {\rm MeV})=2.34 \times 10^{-8}f_{_{\rm G}}
 \left(\frac{M_{\rm gas}}{10^8 \msol}
\right) \left(\frac{d}{100 \ {\rm kpc}}\right)^{-2} \, \mbox{photons 
cm$^{-2}$ s $^{-1}$} \,. 
\end{equation}
The largest fraction of the gas present in galaxies is in the form of neutral 
hydrogen which is detected via 21cm \ion{H}{1} observations. The
integrated \ion{H}{1}
mass in a galaxy is related to the  integrated
\ion{H}{1} flux $\int S_vdv $ by the scaling
$M_{{\rm H \; I}} \propto d^2$.
This leads to the fortunate circumstance that,
although the calculated gas mass for a 
given galaxy depends on the assumed distance to that galaxy, the ratio
\[\Sigma=\frac{M_{\rm gas}}{d^2}\]
is associated only with quantities which are directly observable 
and is independent of any assumption on the distance. 
Thus, eq. (\ref{flux}) can finally be re-written to express
the $\gamma$-ray flux of photons $> 100$ MeV from galaxy $G$
\begin{equation}\label{newflux}
F_{\gamma}^G=2.34 \times 10^{-8}f_{_{\rm G}}
 \left(\frac{\Sigma}{10^4 \msol {\rm kpc}^{-2}}
\right)\, \mbox{photons 
cm$^{-2}$ s $^{-1}$} \,. 
\end{equation}
in terms of the ratio $f_{_{\rm G}}$ of the supernova rate in 
$G$ to that of the Milky Way, and the gas mass-to-distance squared
ratio $\Sigma$.

\section{Data}

Equation (\ref{newflux}) shows that the information we need to calculate
the expected $\gamma$-ray flux from each galaxy is the SN rate of the
galaxy, and the ratio $\Sigma$.
A summary of these data and the relevant references are presented in Table 1. 
The indicated ranges are the span in the published observational data
and are not representative of the error of each measurement as
estimated by the corresponding authors.
This should give a sense of the systematic uncertainties, but one should 
bear in mind that the overall error could be larger, particularly for 
the supernova
rates. 
The value quoted in the $\Sigma$ column is the mean value of all the 
measurements found in the indicated references while the error is just the
square root of the sample variance.

\begin{table}[tb]

\begin{center}
\caption{
Observed Properties of Selected Local Group Galaxies
}
\smallskip
\begin{tabular}{l|lc|ccc}  \hline \hline
  & SN rate  & Adopted 
        &  \multicolumn{3}{|c}{$\Sigma$ \ \ (10$^4$ \msol\ kpc $^{-2}$)} \\ 
 Galaxy &  (century$^{-1}$)  &  $f$ 
  &  \ion{H}{1}  & ${\rm H_2}$ &  Total \\
\hline
LMC     &  0.1$^{(2)}$, 0.23 $^{(3)}$,0.49$^{(4)}$
        & 0.14
                                              &$22 \pm 6 ^{(1),(11),
                                                (12),(13) }$ & $4.63 ^{(13)}$
                                                & 26.6\\
SMC     &  0.065$^{(3)}$, 0.12$^{(4)}$                  
        & 0.04
        &$17 \pm 4 ^{(1),(5)}$  & $0.76 ^{(13)}$  & $17.8$  \\
M31     &  0.9 $^{(9)}$, 1.21$^{(4)}$, 1.25 $^{(7)}$                  
        & 0.45
        &$0.9 \pm 0.2 ^{(1),(6)}$   & $0.06 ^{(16)}$  & $0.92$  \\
M33     & 0.28$^{(8)}$, 0.35$^{(9)}$, 0.68$ ^{(4)}$                   
        & 0.17
        &$0.26 \pm 0.05 ^{(1)}$   & $0.004 ^{(17)}$  & $0.264$  \\
NGC6822 & 0.04$^{(10)}$                   
        & 0.02
        &$0.05 \pm 0.02 ^{(1)}$   & $0.006 ^{(18)}$  & $0.056$  \\
IC10    & 0.082-0.11$^{(14)}$           
        & 0.04
        &$0.016 \pm 0.003 ^{(15)}$   & $\ga 10^{-5 \,\, (19)}$  & $0.016$  
\\
\hline \hline
\end{tabular}
\end{center}
{\small References: (1) Huchtmeier \& Richter \cite{hbook}, 
  references therein; 
(2) Chu \& Kennicutt \cite{chu};
(3) Kennicutt \& Hodge\cite{kennicutt};
(4) Tammann \etal \cite{tammann};
(5) Stanimirovi\'{c} \etal \cite{stanim};
(6) Braun \& Walterbos \cite{braun};
(7) Braun \& Walterbos \cite{braunm31};
(8) Gordon \etal \cite{gordon};
(9) Berkhuijsen \cite{berkh};
(10) Timmes \& Woosley \cite{timmes};
(11) Luks \& Rohlfs \cite{luks};
(12) Kim \etal \cite{kim};
(13) Westerlund \cite{westerlund};
(14) Thronson \etal \cite{thronson};
(15) Shostak \& Skillman \cite{shostak}, also references therein
(16) Dame \etal \cite{dame};
(17) Wilson \& Scoville \cite{wilson1};
(18) Israel \cite{israel};
(19) Wilson \& Reid \cite{wilson2}
}
\end{table}

In order to calculate $f_{_G}$ 
we also need the SN
rate of the MW. Different authors have produced results which cover a range
of roughly an order of magnitude, depending on the method of calculation. 
The three main methods used are extragalactic SN discoveries, 
SN-related Galactic data relating to massive star formation, chemical 
evolution, and nuclear $\gamma$-ray lines,
and analysis  of the historical record of Galactic SN explosions.
Dragicevich \etal \cite{dragicevich} critically surveyed $\snr{MW}$ 
determinations by different methods. The different results quoted in that
 work are given in Figure 1.  

As noted by Dragicevich \etal,
the Galactic supernova rates estimated using data from 
historical supernovae tend to be 
higher than those based on extragalactic or 
nuclear/$\gamma$-ray line constraints.
This discrepancy might arise if our location
in the Galaxy is ``special''--e.g., near a spiral arm where
star formation is enhanced.  In addition, 
Hatano, Fisher, \& Branch \cite{hfb} suggest that
a large fraction ($\sim 50\%$) supernovae are ``ultradim'' 
($M_V \sim -13$), possibly due to localized shrouding effects.  
Such a population would be missed in extragalactic surveys,
but would appear in the historical record,
and would still contribute to nucleosynthesis and accelerate
cosmic rays.
In our study, we will use extragalactic supernova data
for Local Group objects,
and thus the lower estimates of the Galactic supernova rate
are the appropriate ones to use.
As a ``best bet'' we will adopt 
Dragicevich \etal's \cite{dragicevich}
recommended value of 2.5 SN per century.

\section{Results}

We now combine eq. (\ref{newflux}) with the data presented in Table 
1 to predict $\gamma-$ray flux levels for photons with energies above 100 MeV
originating in the interaction between cosmic rays and interstellar medium
in galaxies of the Local Group. We will use the Magellanic Clouds to 
verify the applicability of our method, since for these systems we can compare 
our predictions with both 
EGRET observational results (detection for the LMC, upper limit for the SMC) 
and other predictions using models based on 
entirely different assumptions: dynamic equilibrium for the
LMC (Fichtel \etal 1991), constant cosmic ray proton-to-electron ratio
for the SMC (Sreekumar \& Fichtel 1991).

We will then use our model to proceed to predictions of $\gamma-$ray fluxes
in galaxies of the Local Group for which there have been either
no previous studies or, in the case of M31, only partial treatments.
For the latter, \"{O}zel \& Berkhuijsen \cite{ozel}
based on the gas content alone arrived at a result in which
the cosmic ray flux scaling between the Milky Way and M31
was treated as a free multiplicative parameter, left to be decided
observationally.

Our predictions, and their implications for GLAST, are summarized
in Table 2, with detailed discussion of each galaxy appearing 
below.
In Table 2, all values refer to $\gamma$-rays $> 100$ MeV. The ``GLAST
Significance'' column refers to the formal
significance expected to be achieved
after a 2-year (nominal GLAST duty cycle) and 10-year (GLAST
lifetime goal) all-sky survey. 
The ``On-Target 5 $\sigma$ Exposure Time'' column refers to the total exposure 
{\em of the object} needed to achieve a 5 $\sigma$ detection. When GLAST is
operating in the normal sky-scanning mode, each individual source is 
in the field of view for only $\sim 20\%$ of the time for each duty cycle,
so the GLAST operation time required to achieve a detection of the same 
significance is typically 5-6 times the on-target exposure time quoted 
(assuming a field of view for GLAST between 2 and 2.4 sr). 
All significances and
exposure times were calculated according to the GLAST specifications 
as described in De Angelis \cite{deangelis} and in the GLAST Science
Requirements Document ({\tt \url{http://glast.gsfc.nasa.gov/science/aosrd}}). 
We have used an effective collector area  of $8000 \ {\rm
cm ^{2}}$ (GLAST requirement, $0.8 \times$ GLAST goal), and taken into
account the (Galactic and extragalactic) $\gamma-$ray background noise
levels as measured by EGRET for the Galactic coordinates of each object. 
Whenever the angular extent of an object in the sky (as derived from 21 cm
maps) exceeded the 1-photon angular resolution of GLAST, the actual size
of the object was used as the relevant solid angle for the collection of 
background noise. The same calculation, when applied to a point source of
flux $2 \times 10^{-9} {\rm photons \,\, cm^{-2} \, s^{-1}}$ located at 
high Galactic latitude, predicted a 5-$\sigma$ detection after a 2-year 
all-sky survey, in accordance to the sensitivity derived in the
GLAST Science Requirements Document.

\subsection{Magellanic Clouds}

\subsubsection{Large Magellanic Cloud}

The LMC is the only galaxy other than the Milky Way for which
there has been a positive detection of its diffuse $\gamma-$ray
emission, and is therefore the only one of the systems of interest
for which any prediction can be directly tested against observations. 

For the LMC, Table 1 suggests a mean $\Sigma$ equal to $26.6 \times 10^4
{\rm \msol \, kpc^{-2}}$
and a mean SN rate of 0.27 ${\rm century ^{-1}}$ 
which, combined with a Galactic SN rate of 2.5 $\rm{century ^{-1}}$,
gives 
$f_{{\rm LMC}}=0.11$. Inserting these data in eq. (\ref{newflux}) we
derive  a $\gamma$-ray 
flux for photons with energies $> 100 \ {\rm MeV}$ of 
$6.8 \times 10^{-8}$ photons ${\rm cm ^{-2} \, s^{-1}}$. 

\begin{table}[tb]
\begin{center}
\caption{Predicted Gamma-Ray Flux and GLAST Requirements for 
Selected Local Group Galaxies}
\label{tab:predict}
\begin{tabular}{l|rc|cc|c}
\hline\hline
  & \multicolumn{2}{|c|}{Flux $> 100$ MeV 
     (${\rm photons \; cm^{-2} \; s^{-1}}$)}
  & \multicolumn{2}{|c|}{GLAST Significance}
  & GLAST On-Target  \\ \cline{2-3} \cline{4-5}
Galaxy  & Prediction  & EGRET Value/Limit  
  & 2 years & 10 years 
  & $5\sigma$ Exposure Time \\
\hline
LMC  & $\ee{11}{-8}$  & $\ee{(14.4 \pm 4.7)}{-8}$  
  & $42 \ \sigma$  & $93 \ \sigma$  &  $4.6\times 10^{-3}$ yr  \\
SMC &  $\ee{1.7}{-8}$  &  $< \ee{4}{-8}$
  & $19 \ \sigma$  & $43 \ \sigma$  & $2.1\times 10^{-2}$ yr  \\
M31 &  $\ee{1.0}{-8}$  &  $< \ee{1.6}{-8}$
  & $13 \ \sigma$  & $31 \ \sigma$  &  $4.1\times 10^{-2}$ yr  \\ 
M33 &  $\ee{0.11}{-8}$  &  N/A 
  & $1.9 \ \sigma$  & $4.1 \ \sigma$  &  $2.31$ yr  \\
NGC6822 &  $\ee{2.6}{-11}$  &  N/A 
  & $0.04 \ \sigma$  & $0.09 \ \sigma$  &  $\gg 10$ yr  \\
IC10 &  $\ee{2.1}{-11}$  &  N/A 
  & $0.02 \ \sigma$  & $0.05 \ \sigma$  &  $\gg 10$ yr  \\
\hline\hline
\end{tabular}
\end{center}
{\small 
\noindent
}
\end{table}

However, of the 3 references quoted in Table 
1 for the SN rate of the LMC, the lowest one
 (ref. (2) in the table, equal to 0.1 SN 
${\rm century^{-1}}$), which is based on a count of 
observed SN remnants,  is derived only as a lower 
limit to the LMC SN rate.
The other two estimates are based on extragalactic
SN discoveries in morphologically 
similar galaxies (ref. 4 in Table 1) and on the 
massive star formation rate (ref. 3 in Table 1). If 
we use the mean value of the 
latter two as our best estimate for $\snr{LMC}$
we get $f_{{\rm LMC}}=0.14$.

As far as the gas mass is concerned, 
although the more recent 21cm surveys
tend to give rather low values for $\Sigma$ 
(Luks \& Rohlfs 1992, Kim \etal 1998), the gas mass estimates in 
those cases are assuming an optically thin medium. However, 
recent studies  of the cool gas in the LMC 
by Marx-Zimmer \etal \cite{zimmer} are not in favor of this assumption, 
which indicates that the gas masses might in fact be significantly 
underestimated. On this basis, we will adopt for our calculation the
 higher estimate from Westerlund 
\cite{westerlund} which predicts a $\Sigma_{\mbox{\ion{H}{1}}}=28 \times 10^4
\msol \, {\rm kpc}^{-2}$, and thus 
$\Sigma_{\rm tot} = 32.6 \times 10^4 \msol \, {\rm kpc}^{-2}$.
 
These values of $f$ and $\Sigma$, if used in eq. (\ref{newflux}), 
yield a total 
$\gamma$-ray flux of 
\begin{equation}\label{LMC}
F_{\gamma}^{\rm LMC}=11 \times 
10^{-8} \, 
\mbox{photons cm$^{-2}$ s $^{-1}$} \,. 
\end{equation}
This value is in excellent agreement with
the observed value of $(14.4 \pm 4.7) \times 10^{-8} \mbox{
photons cm$^{-2}$ s $^{-1}$}$
(Hartman \etal 1999).
This consistency gives us confidence in our method
of computing galactic cosmic ray fluxes.

Indeed, one could even turn the argument around, and 
tentatively interpret this agreement as an indication that
the cosmic ray confinement in the LMC is
comparable to that of the Milky Way, $\Lambda_{\rm esc} \sim 10 \ {\rm g \; cm^{-2}}$.
Given the large differences in the size and structure of these
two galaxies, this result would suggest that 
cosmic ray confinement is not strongly dependent on a galaxy's global
properties, but more closely connected to local physics,
such as the magnetic field strength configuration and fluctuations.
If confirmed, therefore, this result would give new and unique information
about the nature of cosmic ray confinement and propagation.
As noted below (\S \ref{sect:m31}), this result can be tested by combining the flux
measurements for multiple Local Group sources, particularly by
comparing the LMC to the Small Magellanic Cloud.

Of course, the uncertainties in the
observational data, especially the Galactic SN rate, can
have an important impact on our calculation.
For example, if we adopt a Galactic SN rate of 1 ${\rm century ^{-1}}$
and keeping the same
values 
for all other data, 
the LMC flux 
could be as large as 27 $\times 10^{-8}  
{\rm photons \,\, cm^{-2} \,s ^{-1}}$ 
On the other hand, if the Galactic SN
rate is as high as 5 ${\rm century ^{-1}}$, this estimate would drop to 5.3 
$\times 10^{-8} \ {\rm photons \,\, cm^{-2} \, s^{-2}}$. 

With these caveats in mind, our best estimate for the
flux gives a very strong detection (formally, at the 42$\sigma$ level)
in the first 2 years of
sky-scanning GLAST operation, with the 5-$\sigma$ detection feasible after 
an on-target observation time of less than 2 days. 

\subsubsection{Small Magellanic Cloud}

The $\gamma-$ray flux level of the Small Magellanic Cloud has 
been the object of both theoretical and observational studies 
in the past, which have been used to resolve the debate of the
origin of the cosmic rays: if the origin of the cosmic rays
were cosmological and the level of the CR flux were universal
the $\gamma-$ray flux should have been easily detectable
by EGRET (Sreekumar \& Fichtel, 1991). However, the SMC was
not detected by EGRET, and an observational
upper limit was placed instead on its $\gamma-$ray flux.
(Sreekumar \etal 1993, Lin \etal 1996). 

This observational upper limit was also lower than 
the value Sreekumar \& Fichtel \cite{sreeksmc} predicted if the 
cosmic rays, gas and magnetic fields in the SMC were in a state of dynamic
equilibrium. Thus, Sreekumar \etal (1993) attributed the non-detection of
the SMC by EGRET to poor confinement of cosmic rays in the SMC. 

In the framework of our model, we will investigate whether CR flux levels of
the SMC low enough so as to prohibit its detection by EGRET can also
be explained in terms of a low supernova rate and consequently a low cosmic
ray acceleration rate, even if we assume similar confinement conditions
with those of the Milky Way. Although the differences between the
morphologies of the MW and the Magellanic Clouds might raise questions
concerning the validity of the latter assumption, the good
agreement of our results with observations in the case of the LMC
encourage the application of our method in the case of the SMC as well. 

For the SMC,
the mean value of $\Sigma$ is equal to $17.8 \times 10^4 {\rm \msol
\,\, kpc^{-2}}$ and the average of the quoted SN rates is 0.09 SN per   
century (Table 1). Thus, eq. \ref{newflux} (using again a $\snr{MW}$ equal to
2.5 $ {\rm century }^{-1}$) predicts a $\gamma$-ray flux of
\begin{equation}
F_{\gamma}^{\rm SMC} = 1.7 \times 10^ {-8} \ {\rm photons \, \, \, cm^{-2}
\,
s^{-1} } \, .
\end{equation}
This value is consistent with the current observational upper limit of
$4 \times 10^{-8} $ photons ${\rm cm^{-2} \, s^{-1} }$ of 
Lin \etal \cite{lin}.
Using this ``best bet'' value for the SMC $\gamma-$ray flux, we find that
GLAST will detect the SMC with a 19 $\sigma$ significance after a 2-year
all-sky survey. The total exposure time needed to achieve a 5 $\sigma$
detection is only 8 days. 

We thus see that the observed SN rate for the SMC is, by itself,
sufficient to explain the observational upper limit placed by EGRET, even 
under the assumption of confinement conditions that do not differ from
those in the Milky Way. Whether confinement in the SMC plays an additional
role in lowering the global cosmic ray and subsequently $\gamma-$ray flux
is a very interesting question, left to be answered observationally by
GLAST.

If now we use the upper limit values for the $\Sigma _{\rm SMC}$ 
and $f_{\rm SMC}$ of $21.8 \times 10^4 {\rm \msol \,\, kpc^{-2}}$ and
$0.05$ correspondingly, the resulting $\gamma-$ray flux reaches $2.6
\times 10^{-8} \ {\rm photons \,
\, \, cm^{-2} \, s^{-1} }$, still below the current observational upper
limit.
In the other end of the range, using the lowest estimates for the 
$\Sigma$ and $\snr{}$ of the SMC, the result we get is equal to
$0.85 \times 10 ^{-8} \ {\rm photons \,\, \, cm^{-2} \, s^{-1} } \,$,  well
above the anticipated
sensitivity of GLAST.
In fact,  the SMC $\gamma-$ray flux would be
still detectable by GLAST even if, in addition to using the lowest
available values for $\Sigma$ and $\snr{SMC}$, we adopted a Galactic SN
rate higher than our ``best bet'' by a factor of 3.

Although our assumption for a ``leaky box'' 
CR confinement is more questionable in
the case of the Magellanic Clouds, our prediction is consistent
the upper limit set by EGRET.
Furthermore, it coincides
with the detailed calculations of Sreekumar \& Fichtel \cite{sreeksmc}.
These authors used synchrotron radiation measurements to deduce the
cosmic ray
intensity and distribution in the SMC (assuming a CR proton-to-electron
ratio same as that of the MW) and predicted a total flux of
$\gamma$-rays above 100 MeV equal to $1.7 \times 10^{-8} \ {\rm photons \,
\, \, cm^{-2} \, s^{-1} } \, .$  
The perfect agreement between this prediction and ours is fortuitous,
but nevertheless increases our confidence in our basic model.
In the cases of galaxies such as M31 and
M33, we expect the CR confinement conditions to be much more similar to
those in the MW (observations indicating the opposite would not only be
a surprising but a very interesting result in  itself). Thus, we expect
our predictions for these galaxies to be, within our
uncertainty limitations, even more reliable.

\subsection{M31}
\label{sect:m31}

The mean observed value of $\Sigma$ for M31 is $0.92 \times 10^4
{\rm \msol \,\, kpc^{-2}}$ while the average of $\snr{M31}$ as
measured
with all 3 different methods (observations of SN remnants, star formation
rates and morphology arguments) is 1.12 per century (Table 1).
The average $\snr{M31}$ corresponds to an M31/MW supernova rate
ratio $f_{{\rm M31}}=0.45$. 
It is worth pointing out that
M31 has an unusually low H$\alpha$ and far infrared emission
(e.g., Pagini et al.\ cite{pagnini}),
which imply a low star formation rate, and hence 
a low Type II supernova rate for such a large
galaxy.  
As we will see, GLAST should detect M31,
an thus provide an important new measure of the M31 supernova rate.

Using our adopted gas content and supernova rate for
M31, eq. (\ref{newflux}) then predicts a total
$\gamma$-ray flux for energies above 100 MeV
\begin{equation}
F_{\gamma}^{\rm M31}=1.0 \times
10^{-8} \,
\mbox{photons cm$^{-2}$ s $^{-1}$} \,.
\end{equation}
This value is consistent with the observational upper limit of
$1.6 \times 10^{-8} \ {\rm photons \,\, cm^{-2} \,\, s^{-1}}$   
set by Blom \etal \cite{blom}, but only slightly lower than the EGRET
sensitivity. 
As we see in Table 2, our predicted $\gamma-$ray flux for M31 would be
detected by GLAST in its first 2-year all-sky survey with a 14 $\sigma$
significance. After a projected lifetime of 10 years, and assuming
continuous sky-scanning operation, this significance
would rise to 31 $\sigma$.  

Our calculation gives theoretical support to the case made
by Digel et al.\ \cite{dmosw}, who noted that
if the flux from Andromeda lies just below the Blom et al.\ limit,
then GLAST will readily be able to detect it.  
Here we find that indeed, the flux should be just at the level
of $1 \times 10^{-8} \, \mbox{photons cm$^{-2}$ s $^{-1}$}$
suggested by Digel et al.,
though the uncertainty range is considerable,
as we now see.

To estimate the effect of the various uncertainties in our
calculation, we observe that if we had
used the highest available observed values for $\Sigma$ and $\snr{M31}$
as given in Table 1 but kept the Galactic SN rate fixed at 2.5 per  
century, the resulting flux of $1.3 \times 10^{-8} \ {\rm photons \,\,
cm^{-2} \,\, s^{-1}}$ would still be lower than the observational upper
limit. With $\Sigma$ and $\snr{M31}$ values both at the upper end of their
respective ranges, $F_\gamma^{\rm M31}$ would remain below the EGRET
limit for $\snr{MW}$ as low as 2 per century. 
 
At the other end of the range of the available M31 data, using a $\Sigma$
as low as $0.7 \times 10^4 {\rm \msol \,\, kpc^{-2}}$ and $\snr{M31}$
of 0.9
per century, we derive a predicted flux of $ 0.6 \times 10^{-8}\ {\rm
photons \,\, cm^{-2} \,\, s^{-1}}$, still easily above the expected GLAST
sensitivity. If in addition we take into account the uncertainty in the
$\snr{MW}$, we can see that even with the most pessimistic estimates for  
 the M31 gas content and SN rate, the $\gamma$-ray flux remains above the
GLAST detectability limit for a Galactic SN rate up to 7 per century.   

In sum, our analysis leads to the robust prediction that a $\gamma$-ray
observatory with sensitivity similar to that expected for GLAST will
detect the diffuse $\gamma-$ray signature of M31.

Once there is a positive detection, depending on the strength of the   
signal and the available spatial resolution, this opens 
several avenues of further analysis:
\begin{itemize}
\item The spatial distribution of neutral hydrogen in M31 has been
observed to exhibit an asymmetry, with the west part of the galaxy
having a significantly lower column density than the northeast part.
For the relevant $\gamma$-ray fluxes of the two areas (a few times
$10^{-9}$ up to 
$10^{-8}$ ${\rm photons \,cm^{-2}\,s ^{-1}}$), the angular resolution 
of GLAST will be $10^\prime - 
20 ^\prime$ (Digel \etal 2000). Given the $\sim 2^\circ$ size
of 
M31 in 21 cm (which is the relevant angular
size for $\gamma-$ray emission), this asymmetry is easily within the
angular resolution capabilities of GLAST 
and is thus expected to be observed in the  
$\gamma-$ray signal as well.
If the M31 flux is high enough, one might even hope to observe
effects of the magnetic torus (e.g., Beck, Brandenburg, Moss, Shukurov,
\& Sokoloff \cite{bbmss}) and star forming ring
(e.g., Pagani et al.\ \cite{pagani}) 
at radius 10 kpc. A similar morphological feature in the 
Milky Way, the ${\rm H_2}$ ring extending in radius from 4 to 8 
kpc (e.g., Bronfman \etal 1988) was first detected in early $\gamma-$ray 
surveys and then shown to be consistent with
subsequent CO molecular surveys 
(Stecker, Solomon, Scoville, \& Ryter 1975).

\item An observational measurement of the diffuse $\gamma-$ray flux from
M31 could be used to reverse the arguments presented in section
\ref{sec:theory} and make inferences regarding the cosmic ray flux in M31:  
From the $\gamma-$ray flux as measured from Earth, 
$F_{\gamma}^{\rm M31}$, and using the distance and gas content of M31, we
can calculate the $\gamma$-ray production rate per target H-atom. This
result, combined with an assumption for the cosmic ray energy
spectrum, would provide information on the cosmic ray flux level
in M31 which could then be compared with the MW cosmic ray flux to
determine whether they are comparable and whether the MW has a typical
cosmic ray activity for galaxies with similar morphology.
(Note here that we assume that the contribution from
point sources in M31 would be
negligible.)

\item A measured value of the $\gamma-$ray flux from M31 could 
also be used to determine the ratio of the
supernova rates between M31 and the
Milky Way, assuming that the gas content of M31 is known to a good
accuracy:
\beq
f_{_{\rm M31}} = \frac{\snr{M31}}{\snr{MW}}
=0.45 \left(\frac{F_\gamma^{\rm M31}}{10^{-8} \,{\rm photons \,
cm^{-2}\,s^{-1}}}\right)
\eeq
using eq. \ref{newflux}.
 This ratio could then be used e.g. to determine the MW supernova
rate or to to infer the ratio of formation
rates for high-mass stars, and, assuming a constant initial mass function,
the ratio of total star formation rates between the two galaxies. 
(Here we rely on our assumption that all cosmic rays are produced
by supernovae.)
  
\end{itemize}

These last two points can be self-consistently checked once
GLAST has information on more than one extragalactic source.
For example, the ratios of the flux measurements {\em among} the
extragalactic sources is a direct measure of the ratios of the cosmic ray
densities.  In our model, these ratios gives the ratio of the 
product of $\snr{} \lesc$.
Thus, with information about the supernova rates, one can directly measure
the ratios of cosmic ray confinements; this allows one to compare
these values between the LMC and SMC, and to contrast the clouds with
M31 (and M33).  
One might then assume, in the case of the clouds, that their confinements are equal, and
compare the inferred supernova rates with values
estimated by other means.
Alternatively, one could assume (as we have) that the confinements
are these same as in the MW, and then use
the ensemble of measurements of Local Group sources to
reduce the uncertainty in the inferred Milky Way SN rate.

Another potential cross-check would come from the comparison of
Local Group $>100$ MeV emission and $\gamma$-ray line emission from
radioactive decays (Timmes \& Woosley \cite{timmes}). 
The line flux is sensitive to the supernova rate only, so that
the ratio of the $>100$ MeV continuum to the lines gives 
direct information about the cosmic ray confinement.
Unfortunately, such a comparison may have to wait, as 
the Timmes \& Woosley calculations predict that the 
line fluxes lie below the expected sensitivity of the
forthcoming
{\em International Gamma-Ray Astrophysics Laboratory} (INTEGRAL)
mission.

Even a non-detection of M31 in high energy $\gamma$-rays would be, apart from
very surprising, a result of high theoretical
interest, since the only parameter entering our calculations other that
$f$ and $\Sigma$ is $\lesc$ (the escape mean free path)--which in turn
depends on the confinement conditions of the cosmic ray nuclei in M31. A
non-detection would thus lead us to reconsider our assumption of similar
confinement conditions with M31. Such a result would provide
observational clues for the confinement in an extragalactic
environment (morphologically similar to the Milky Way) 
and for the nucleonic component of cosmic rays.
 
\subsection{Other Local Group Galaxies}
 
Using the mean values for the gas content-related $\Sigma$ and the
supernova rate shown in Table 1 for the case of M33, eq. (\ref{newflux})
gives a $\gamma-$ray flux of
\begin{equation}
\label{eq:m33}
F_\gamma^{\rm M33} = 0.11 \times 10^{-8} \ {\rm
photons \,\, cm^{-2} \,\, s^{-1}} \,,
\end{equation}
which is slightly below the sensitivity limit of GLAST. 
As seen in Table 2, after a 10 year sky survey, 
the detection is at the $4.2\sigma$ level (comparable to the current
LMC significance).
If we use 
the highest available estimates for the M33 gas content and supernova   
rate, the flux would rise to $0.2 \times 10^{-8}\  {\rm
photons \,\, cm^{-2} \,\, s^{-1}}$, which would allow for
a $5\sigma$ detection after a sky survey of 4.39 years. 
Alternatively, even if eq.\ \pref{eq:m33} is accurate, 
M33 will be detectable at the $5\sigma$ 
level within 10 years 
if the GLAST effective area and field of view achieve their 
``goal'' levels (as opposed to the ``required'' levels we have used).

Thus, there is grounds for
optimism that a detection might be possible, if either the gas content and
supernova rate have been slightly underestimated, or the actual sensitivity
of the  GLAST observatory
is slightly improved as compared to the current prediction.
We therefore urge observers to be aware of the possibility of
detecting (or placing a limit on) flux from M33.
 
The next best candidates for detection (highest combination of $\Sigma$   
and $f$) in the local group are NGC6822 and IC10. However, using the data
shown in Table 1 for these galaxies, eq. (\ref{newflux}) predicts fluxes 
which are comparable within our uncertainty limits and equal to about
$0.002 \times 10^{-8} \ {\rm photons \,\, cm^{-2} \,\, s^{-1}}$. Such 
tiny fluxes would require exposure GLAST times of decades,
and thus appear to lie beyond reach for the foreseeable future.
 
Thus, apart from the Magellanic Clouds, M31 and M33, no other Local Group
galaxy seems to be a good candidate for the detection of its diffuse    
$\gamma$-ray flux by $\gamma$-ray observatories in the near future. 
There might,
however, be other promising candidates, which lie outside the Local Group. Starburst
galaxies, despite the fact that they lie at a greater distance that Local 
Group galaxies, have a significantly higher supernova rate, as well as
high gas contents, which should result to high $\gamma$-ray production rate
and a flux that might be detectable by highly sensitive observatories
(Paglione \etal 1996).

\section{Discussion}

Diffuse, high-energy ($\ga 100$ MeV) 
$\gamma$-rays
provide the most direct evidence for the extension of the cosmic ray
ion component throughout our Galaxy.
The observation of such radiation from extragalactic
systems would provide unique information,
as the mere detection of high-energy 
$\gamma$-rays confirms the presence of cosmic ray
ions, and the photon flux can be used to infer the 
cosmic ray flux, and thus can constrain 
extragalactic cosmic ray properties.
Unfortunately, the only extragalactic object
detected thus far is the LMC (Sreekumar {\it et al} \cite{sreeklmc}).

In anticipation of future high-energy
$\gamma$-ray observatories such as GLAST,
we have estimated the $\gamma$-ray flux
due to diffuse emission for Local Group galaxies.
To do this, we have used a simple ``leaky box'' model
of cosmic ray propagation, and taken
supernova blasts to be the engines of cosmic ray acceleration.
Our model makes different assumptions than
Fichtel \etal's \cite{fichtelmc} 
more detailed treatment of the LMC,
but both give similar results, and are in
reasonable agreement with LMC $\gamma$-ray observations.

Applying our model to other Local Group galaxies, we
predict that M31 has a $\gamma$-ray flux above
100 MeV of about 
$1.0\times 10^ {-8}\  {\rm photons \, \, \, cm^{-2} \, s^{-1}}$, 
with an uncertainty of about
a factor of 3.  Fortunately, despite this large error budget,
we can conclude that M31 should be observable by GLAST,
and we therefore strongly urge that M31 be looked for in 
GLAST maps.
A detection will provide important and unique information about
cosmic rays in a galaxy similar to our own.
In addition, we find that the SMC should 
have a flux of about 
$ 1.4 \times 10^ {-8} \ {\rm photons \, \, \, cm^{-2} \, s^{-1}}$, 
readily detectable by GLAST.
The comparison among the LMC, SMC, and M31 $\gamma$-ray luminosities
will provide new information about cosmic ray densities and confinement,
and supernova rates, in these systems and in the Milky Way.

The high-energy $\gamma$-ray flux from other Local Group galaxies
is much smaller.  Other than M31 and the Magellanic
clouds, the only system that is potentially observable
is M33, with a flux of about 
$0.1 \times 10^ {-8} \ {\rm photons \, \, \, cm^{-2} \, s^{-1}}$. 
If GLAST can stretch to reach its sensitivity goals, this
too will be observable.
All other Local Group galaxies have emission that is 
at least 2 orders of magnitude smaller.

\acknowledgments
We thank David Branch,
Jim Buckley, You-Hua Chu, Robert Gruendl and Kostas Tassis for
enlightening
discussions.
We thank the referee for constructive comments  which
have improved this paper.
The work of V.P.  was partially supported by a scholarship
from the Greek State Scholarship Foundation.
\nobreak

\newpage

\begin{figure}[t]
\epsscale{0.8}
\plotone{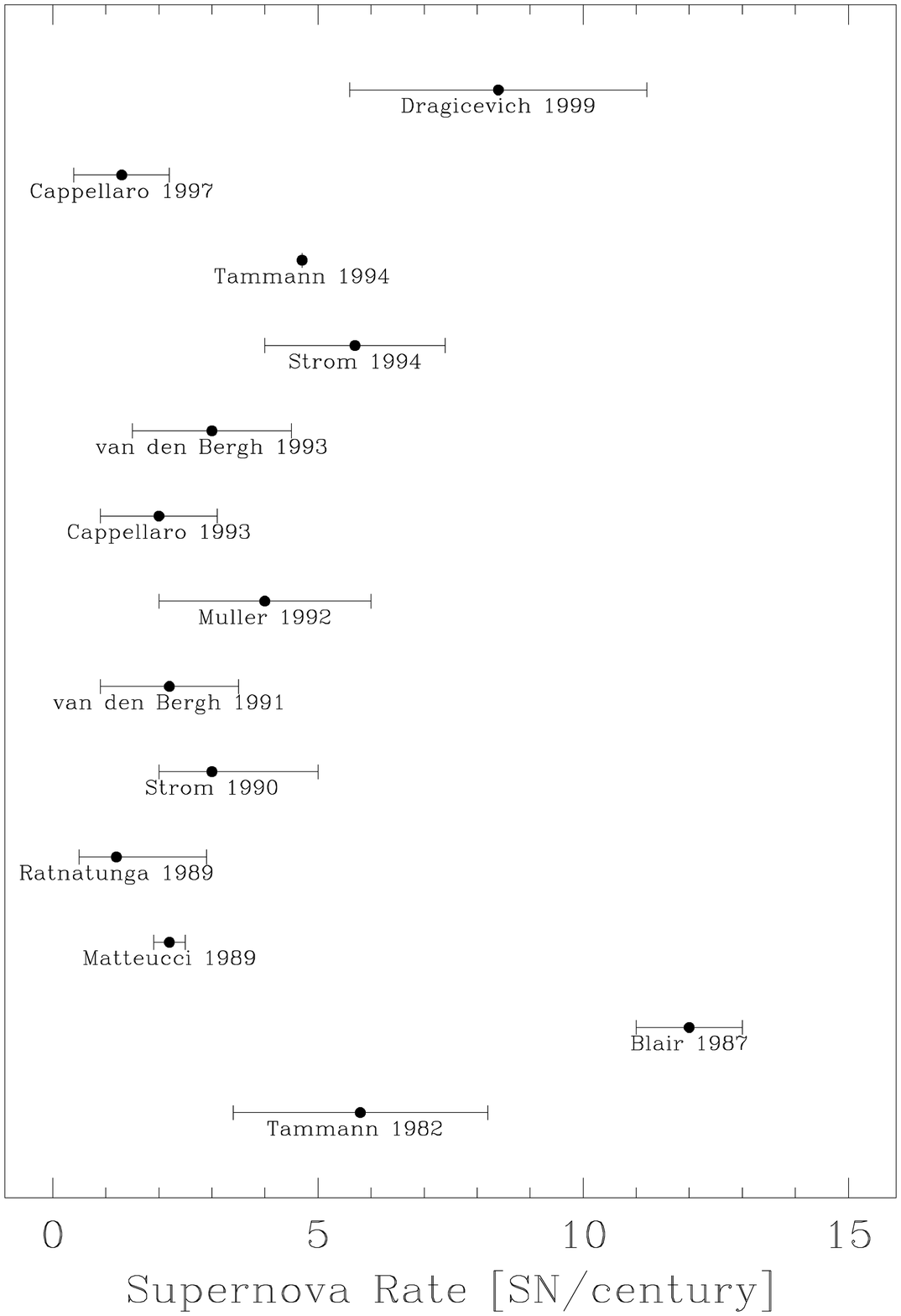}
\caption{Milky Way supernova rates as calculated by different authors,
from the tabulation of Dragicevich \etal \cite{dragicevich}. The
estimates from
extragalactic SN discoveries have been standardized to 
$h=0.75$ and a 
Milky Way blue luminosity of $(2.3 \pm 0.6) \times 10^{10} L_{\odot,B}$.}
\end{figure}


\begin{thebibliography}{blom}

\bibitem[1991]{akyuz} Aky\"{u}z, A., Brouillet, N. \& \"{O}zel, M.E. 1991, 
A\&A, 248, 419

\bibitem[1996]{bbmss} Beck, R., Brandenburg, A., Moss, D. Sukurov, A., 
\& Sokoloff, D. 1996, ARAA, 34, 155

\bibitem[1990]{bbdgp} Berezinski\u{\i}, V.S., Bulanov, S.V., Dogiel, V.A.,
Ginzburg, V.L., \& Ptuskin, V.S. 1990, Astrophysics of Cosmic Rays,
 Amsterdam: North-Holland

\bibitem[1984]{berkh} Berkhuijsen, E.M. 1984, A \&A, 140, 431

\bibitem[1999]{blom} Blom, J. J., Paglione, T. A. D. and
Carrami{\~n}ana, A.  1999, ApJ, 516, 744 

\bibitem[1992]{braun} Braun, R. \& Walterbos, R.A.M. 1992, ApJ, 386, 120

\bibitem[1993]{braunm31} Braun, R. \& Walterbos, R.A.M. 1993, A\&A, 98, 327

\bibitem[1988]{bron} Bronfman, L., Cohen,
R. S., Alvarez, H., May, J., \& Thaddeus, P. 1988, ApJ, 324, 248

\bibitem[1971]{cavallo} Cavallo, G. \& Gould, R.J. 1971, Nuovo Cimento,
2B, 77

\bibitem[1988]{chu} Chu, Y.-H. \& Kennicutt, R.C. Jr. 1988, ApJ, 96, 6

\bibitem[1998]{combi} Combi, J.A., Romero, G.E. \& Benaglia, P. 1998, 
A\&A, 333, L91

\bibitem[1980]{cram} Cram, T.R., Roberts, M.S. \& Whitehurst, R.N. 1980, 
A\&AS, 40, 215

\bibitem[1993]{dame} Dame, T. M., Koper, E., Israel, F. P. \& Thaddeus, P.
1993, ApJ 418, 730

\bibitem[1986]{dermer} Dermer, C. D. 1986, A\&A, 157, 223

\bibitem[2000]{deangelis} De Angelis, A. 2000, to be published in the 
Proceedings of the 3rd International Workshop ``New Worlds in
Astroparticle Physics,'' astro-ph/0009271 

\bibitem[1995]{digel} Digel, S.W., Hunter, S.D. \& Mukherjee, R. 1995,
ApJ, 441, 270

\bibitem[2000]{dmosw} Digel., S.W., Moskalenko, I.V., 
Ormes, J.F., Sreekumar, P., \& Williamson, P.R.
2000, AIP Conf.Proc., 528, 449

\bibitem[1999]{dragicevich} Dragicevich, P.M., Blair, D.G. \& Burman,
R.R. 1999, MNRAS, 302, 693 

\bibitem[1997]{edm} Ellison, D.C., Drury, L.O., \& Meyer, J.P.
1997, ApJ, 487, 197

\bibitem[1984]{fichtel1} Fichtel, C.E. \& Kniffen, D.A., A\&A, 134, 13

\bibitem[1991]{fichtelmc} Fichtel, C.E., \"{O}zel, M.E., Stone, R.G. \&
Sreekumar, P. 1991, ApJ, 374, 134 

\bibitem[1998]{gordon} Gordon, S.M., Kirshner, R.P.,
 Long, K. S., Blair, W. P., Duric, N. \& Smith, R.C. 1998, ApJS, 117, 89

\bibitem[1999]{hartman} Hartman R.C. \etal 1999, ApJS, 123, 79

\bibitem[1997]{hfb} Hatano, K., Fisher, A., \& Branch, D. 1997, MNRAS, 290, 360

\bibitem[1973]{huchtmeier} Huchtmeier, W. K. 1973, A\&A, 22, 91 

\bibitem[1989]{hbook}Huchtmeier, W. K., \& Richter, O. 1989, 
A general catalog of \ion{H}{1} 
observations of galaxies: the reference catalog, (Springer-Verlag: New York)

\bibitem[2000]{hucht} Huchtmeier, W. K, Karachentsev, I.D., Karachentseva, 
V.E. \& Ehle, M. 2000, A\&A 141, 469

\bibitem[1997]{hunter} Hunter, S.D. {\it et al} 1997, ApJ 481, 205

\bibitem[1997]{israel} Israel, F. P., 1997, A\&A 317, 65

\bibitem[1986]{kennicutt}Kennicutt, R. C., Jr. \& Hodge, P. W. 1986, ApJ,
 306,130

\bibitem[1998]{kim} Kim, S., Staveley-Smith, L., Dopita, M.A., Freeman, K.C., 
Sault, R.J., Kesteven, M.J. \& McConnell, D. 1998, ApJ, 503, 674

\bibitem[1995]{koyama} Koyama, K. \etal 1995, Nature, 378, 255

\bibitem[1996]{lin} Lin, Y.C. \etal 1996, ApJS, 105, 331

\bibitem[1990]{long} Long, K.S., Blair, W.P., Kirshner, R.P. \& Winkler, P.F.
1990, ApJS, 72, 61

\bibitem[1992]{luks} Luks, Th. \& Rohlfs, K. 1992, A \& A, 263, 41

\bibitem[2000]{zimmer} Marx-Zimmer, M, Herbstmeier, U. Dickey, J.M., 
Zimmer, E., Staveley-Smith, L. \& Mebold, U. 2000, A \& A, 354, 787

\bibitem[1997]{mori} Mori, M. 1997, ApJ, 478, 225

\bibitem[1987]{ozel} \"{O}zel, M.E. \& Berkhuijsen, E.M. 1987, A\&A, 172,
378

\bibitem[1988]{ozfichtel} \"{O}zel, M.E. \& Fichtel, C.E. 1988, ApJ 335,
135

\bibitem[1999]{pagani} Pagani, L., et al.\ 1999, A\&A, 351, 447 

\bibitem[1996]{paglione} Paglione, T.A.D., Marscher, A.P., Jackson, J.M. \&
Bertsch, D.L. 1996, ApJ, 460, 295

\bibitem[1989]{shostak} Shostak, G.S. \& Skillman, E.D. 1989, A \& A, 214,
33

\bibitem[1991]{sreeksmc} Sreekumar, P. \& Fichtel, C.E. 1991, A\&A, 251,
447

\bibitem[1992]{sreeklmc} Sreekumar, P. {\it et al} 1992, ApJ, 400, L67

\bibitem[1993]{sreeksmcobs} Sreekumar, P. {\it et al} 1993, Phys.Rev.L.,
70, 127

\bibitem[1998]{sreekumar} Sreekumar, P. {\it et al} 1998, ApJ, 494, 523 

\bibitem[1999]{stanim} Stanimirovi\'{c}, S.,Staveley-Smith, L., Dickey, J.M., 
Sault, R.J. \& Snowden, S.L. 1999, MNRAS, 302, 417 

\bibitem[1970]{stecker1} Stecker, F.W. 1970, Ap. and Space Sci., 6, 377

\bibitem[1973]{stecker2} Stecker, F.W. 1973, ApJ, 185, 499

\bibitem[1975]{stecker4} Stecker, F.W., Solomon, P.M., Scoville, N.Z.
 \& Ryter, C.E. 1975, ApJ, 201, 90 

\bibitem[1988]{stecker3} Stecker, F.W. 1988, in Cosmic Gamma Rays,
Neutrinos and Related Astrophysics, ed. M.M. Shapiro \& J.P. Wefel
(Dordrecht:Reidel), 85

\bibitem[1981]{stephens} Stephens, S.A. \& Badhwar, G.D. 1981, Ap\&SS, 76,
213

\bibitem[1996]{strong} Strong, A.W. 1996, SSRv, 76, 205

\bibitem[2000]{smr} Strong, A.W., Moskalenko, I.V.,
\& Reimer, O. 2000, ApJ, 537, 763

\bibitem[1994]{tammann} Tammann, G.A., L\"{o}ffler, W. \& Schr\"{o}der, A. 1994,
ApJS, 92, 487

\bibitem[1997]{timmes} Timmes, F. X. \& Woosley, S. 1997, ApJ, 489, 160

\bibitem[1990]{thronson} Thronson, H. A. Jr., Hunter, D. A., Casey, S. \&
Harper, D. A. 1990. ApJ, 355, 94

\bibitem[1997]{westerlund} Westerlund, B.E. 1997, The Magellanic Clouds, 
(Cambridge:  Cambridge Univ. Press), 28

\bibitem[1989]{wilson1} Wilson, C.D. \& Scoville, N. 1989, ApJ 347, 743

\bibitem[1991]{wilson2} Wilson, C.D. \& Wilson, Reid, I. N. 1991, ApJL 366, 11

\end{thebibliography}
\end{document}